\documentstyle[prd,aps]{revtex}

\def\beq{\begin{equation}}
\def\eeq{\end{equation}}

\def\lapprox{\lower 3pt\hbox{$\buildrel < \over \sim\;$}}
\def\gapprox{\lower 3pt\hbox{$\buildrel > \over \sim\;$}}

\def\geff{\gamma_{\rm eff}}
\def\trho{\tilde \rho}
\def\tp{\tilde p}
\def\tT{\tilde T}

\newcommand{\be}{\begin{equation}}
\newcommand{\ee}{\end{equation}}

\begin{document}

\draft

\title{Scalar field description of decaying-$\Lambda$ cosmologies}

\author{J. M. F. Maia$^{1,2}$ and J. A. S. Lima$^3$}

\address{1. Instituto de F\'\i sica Te\'orica, UNESP, Rua Pamplona,
        145, 01405-900, S\~ao Paulo, SP, Brazil.}
\address{2. FINPE, Instituto de F\'\i sica, Universidade de S\~ao Paulo,  CP
        66318, 05315-970, S\~ao Paulo, SP Brazil. }
\address{3. Departamento de F\'\i sica, UFRN, C.P. 1641, 59072-970, Natal,
        RN, Brazil.}

\maketitle

\vspace{0.5cm}

\begin{abstract}
 The conditions under which cosmologies driven by time varying
cosmological terms can be described by a scalar field coupled to a
perfect fluid are discussed. An algorithm to reconstruct
potentials dynamically and thermodynamically analogue to given
phenomenological $\Lambda$ models is presented. As a worked example,
the deflationary cosmology, which evolves from a pure de Sitter
vacuum state to a slightly modified FRW cosmology is considered. It is
found that this is an example of nonsingular warm inflation with
an asymptotic exponential potential. Differences with respect to
other scalar field descriptions of decaying vacuum cosmologies are
addressed and possible extensions are indicated..

\end{abstract}

\section{Introduction}

Phenomenological models with a time dependent cosmological
``constant'' $\Lambda(t)$ intend to explain how this term, or
equivalently the vacuum energy density, reached its present value.
Usually inspired by qualitative motivations, such proposals may
indicate suggestive ways for solving the cosmological constant
problem, as for instance, by describing the effective regimes that
should ultimately be provided by fundamental physics (for reviews,
see\cite{Overduin}). From a physical viewpoint, decaying vacuum
models can also be attractive as a basis for a more realistic
cosmology as suggested by the latest observations. These models
are in line with recent measurements of luminosity distance based
on SNe type Ia, which are consistently indicating the possible
existence of an unknown form of energy with negative pressure,
like an effective vacuum component, and presumably responsible for
the present accelerating stage of the Universe
\cite{Accelerating}.

Although incredibly small if compared to common microscopic
scales, the cosmological $\Lambda$-term is expected to contribute
dominantly to the total energy density of the universe. Moreover,
since its present value, $\Lambda_0$, may be a remnant of a
primordial inflationary stage, it seems natural to study
cosmological solutions including a decaying vacuum energy density
which is high enough at very early times (to drive inflation), but
sufficiently small at late times in order to be compatible with
the present observations. A possible source for this effective
cosmological term is provided by a scalar field (the so-called
``quintessence'') which has received a great deal of attention
lately \cite{quintessence}, or still a true decaying vacuum energy
density phenomenologically described by an equation of state
$\rho_v(t) = -p_v (t)$.

In what follows, instead of a new decaying vacuum scenario, it is
discussed how such phenomenological cosmologies can be
interpreted in terms of a classical scalar field decaying into a perfect
fluid. This problem deserves particular attention because if a scalar
field version of a $\Lambda (t)$ model can be implemented, its associated
lagrangian can be used in other gravitation theories than general
relativity. This is probably necessary if one wants to search for
fundamental physics formulations of these models by considering
effective high energy regimes in which general relativity is no
longer valid (as in superstring cosmology, for example). Additionally,
the methods employed here are specially addapted for a larger framework
in which the dark energy is coupled to the dark matter, as suggested by
Dalal {\em et al.} \cite{Dalal} (see also \cite{Amendola}). Such
coupled scalar field models may avoid the cosmic coincidence problem, with
the available data being used to fix the corresponding dynamics and,
consequently, the scalar field potential responsible for the present
accelerating phase of the universe. Another interesting feature of coupled
dark energy models is that the temperature dependence on the redshift $z$
of the relic radiation, $T(z)$, can be slightly different from the standard
prediction deduced from the adiabatic FRW type expansion. Indirect
measurements of $T(z)$ at high redshifts may become one of the most
powerful cosmological tests because it may exclude the presence of a
cosmological constant or even any kind of separately conserved quintessence
\cite{LV}.

        This work is organized as follows. In Section II the dynamics
and thermodynamics (along the lines of first order thermodynamics
\cite{Ademir}) of time-varying $\Lambda$ models are reviewed. A
very simple procedure for describing such models in terms of
scalar fields is proposed in section III. This algorithm, the most
important result of this paper, incorporates thermodynamics into
the pioneering treatment of Ellis and Madsen \cite{Ellis}.
Applications of the procedure are discussed in section IV. Special
attention is devoted to the deflationary $\Lambda(t)$ model
suggested for the first time in Refs. \cite{L&M 94}, and
for completeness, some of its variants recently presented in the
literature have also been considered. Section V has a summary of
the results and further applications of the methods used here are
outlined.

\section{Time-varying cosmological term}

        Decaying vacuum cosmologies (see \cite{Overduin}) are described
in terms of a two-fluid mixture: a decaying vacuum medium
($\rho_v=\Lambda(t)/8\pi G,\,\, p_v = -\rho_v$) plus a fluid
component (the decaying vacuum products) described by their energy
density $\rho$ and pressure $p$. For the flat
Friedmann-Robertson-Walker (FRW) line element $(c=1)$, \beq ds^2 =
dt^2-a^2(t)\left(dr^2 + r^2d\theta^2 + r^2\sin^2\theta
d\phi^2\right), \eeq where $a(t)$ is the scale factor. In such a
background, the Einstein field equations (EFE) and the energy
conservation law can be written as

\beq \label{eq:rho4} 8\pi G \rho + \Lambda = 3H^2, \eeq \beq
\label{eq:p4} 8\pi G p - \Lambda = -2\dot H - 3H^2, \eeq \beq
\label{eq:ECL} \dot \rho + 3H(\rho + p) = - {\dot \Lambda \over 8
\pi G}\equiv F, \eeq where $H={\dot a}/a$ is the Hubble parameter
and $F$ denotes a source term for the fluid with energy density
$\rho$ and pressure $p$. Notice that Bianchi identities (or Eq.
(\ref{eq:ECL})) impose that if the cosmological term is a time
decreasing quantity, energy must be transferred from $\Lambda$ to
the perfect fluid. In order to keep the discussion as general as
possible, it is usual to assume that the non-vacuum component
obeys the $\gamma$-law equation of state

\beq p=(\gamma -1)\rho\ , \ \ \ \ \ \ \ \ \ \gamma \in [1,2]\, .
\eeq One may then use (\ref{eq:rho4}) and (\ref{eq:p4}) to find
the useful formula \beq\label{eq:ef} {2 \dot H\over 3H^2} =
-\geff, \eeq where \beq \label{eq:geffL} \geff = \gamma \left(1 -
{\Lambda \over 3H^2}\right).
\eeq
In terms of $\geff$, the
matter-radiation energy density is given by

\beq\label{eq:rhoeflambda} \rho = {3H^2 \over 8\pi G}
{\geff\over\gamma}, \eeq while the source term $F$ appearing in
(\ref{eq:ECL}) can be written as

\beq\label{eq:ECLeff} F = 3H\gamma\rho \left(1 -
{\geff\over\gamma} + {1\over 3H\gamma}
    {\dot\geff\over\geff} \right).
\eeq
The above formulae identify the specific dynamics given for
each different phenomenological $\Lambda(t)$ through the function
$\geff(t)$. As will be seen later, for each phenomenological
expression of $\Lambda(t)$ (see Table 1 of Ref. \cite{Overduin}
for an extensive list of such models and respective references)
the related $\geff$ obtained from (\ref{eq:geffL}) may be
transplanted to the analogue equations in the scalar field
version. If one considers that there is no energy
exchange between the $\phi$ field and the matter or radiation
component, the associated potential can be found (at worst numerically)
\cite{Ellis}. However, this case is not thermodynamically analogue to the
pure $\Lambda(t)$ picture, as it presents particle production. In order to
find the true equivalence, one has to fix the thermodynamics of the
matter/energy creation process and impose the same thermodynamic conditions
to both cases. In order to choose such conditions, it is worth to review
the basics of the first order thermodynamics applied to vacuum decay.

        The approach used here follows the lines set in
Refs. \cite{Ademir,Prigogine}. First of all, it is assumed a
continuous transfer of energy from the decaying vacuum to the
$\gamma$-fluid, as given by Eq. (\ref{eq:ECL}). A more complete fluid
description requires the definition of the particle current $N^{\alpha}$
and the entropy current $S^{\alpha}$ in terms of the fluid variables. If
$n$ denotes the number density of the created particles, the particle
current is $N^{\alpha}=nu^{\alpha}$, and its balance equation can be
written as
\begin{equation}\label{eq:ndot}
{\dot n \over n} + 3H = {\dot N \over N} \equiv \Gamma ,
\end{equation}
where $\Gamma$ is the particle creation rate within a comoving
volume. As a consequence of the vacuum
``equation of state'', the entropy of the mixture depends
exclusively on the matter component. It thus follows that the
entropy current assumes the form
\begin{equation}
S^{\alpha}=n{\sigma}u^{\alpha} ,
\end{equation}
where $\sigma$ is the specific entropy (per particle), and the
mere existence of a nonequilibrium decay process means that
$S^{\alpha}_{{;}\alpha} \geq 0$. Assuming local equilibrium, the
thermodynamic variables are related by Gibbs' law:
\begin{equation} \label{eq:GIBBS}
nTd\sigma = d\rho - {\rho + p \over n}dn ,
\end{equation}
where $T$ is the temperature of the $\gamma$-fluid. From these
relations the temperature evolution law for the fluid component is
given as
\beq \label{eq:EVOLT}
{\dot T \over T} = \left({\partial p \over
\partial \rho}\right)_{n}{\dot n \over n} + {n\dot\sigma \over
\left({\partial \rho \over \partial T}\right)_n}, \eeq where \beq
\dot\sigma = {1\over nT} [F - (\rho + p)\Gamma].
\eeq
Note that if the usual equilibrium relations $n \propto T^{\frac {1}{\gamma -
1}}$, and $\rho \propto T^{\frac {\gamma}{\gamma - 1}}$ are valid,
the specific entropy is constant ($\dot \sigma = 0$) and
\beq  \label{adc}
\gamma\rho\Gamma = F,
\eeq
as previously obtained for
the ``adiabatic'' matter creation process
\cite{Ademir,Prigogine}. This condition is widely used,
specially for the radiation dominated phase, with $\rho \propto
T^4$ and $n\propto T^3$. This is the thermodynamic constraint that
will be fixed for the decaying $\Lambda$ case and will also be
used in its scalar field version\cite{JL}. In particular, this
means that the entropy variation rate of the fluid (in a comoving
volume) is
\beq \label{eq:SrateL}
{\dot S \over S} = {\dot N \over N} =
{F\over \gamma\rho}.
\eeq
It is worth to notice that the relation
(\ref{adc}) between the source of energy and the source of
particles is independent of $\Lambda(t)$, of the Einstein
equations themselves, and it is also valid for the matter
dominated phase. Using (\ref{eq:ECLeff}), (\ref{eq:ndot}) and
(\ref{eq:EVOLT}) one can find the temperature law for the
``adiabatic'' process:
 \beq \label{eq:EVOLTL}
 {\dot T \over T} = -3H{\gamma - 1 \over
\gamma} \left(\geff - {1\over 3H} {\dot\geff\over\geff}\right).
\eeq
A systematic procedure for obtaining scalar field
cosmologies analogue to given $\Lambda(t)$ models is proposed in the next
section.

\section{Scalar field description}

Even when phenomenologically well motivated (as in, for example,
\cite{Darabi}), it is most desirable to have a derivation of time-varying
$\Lambda$ models from fundamental physics. Indeed, there are a few examples
of dynamical $\Lambda$ obtained as a result of fundamental processes. For
instance, in Ref. \cite{Raul}, the back reaction of density perturbations
on the de Sitter spacetime induces a varying vacuum energy density. In
\cite{JEllis}, a relaxing contribution to the cosmological term comes from
string motivated $D$-particle recoil effects. Although promising, these
models are not complete, and some other possibilities might be attempted.
In the former example, only the first stages of a genuine decaying vacuum
cosmology are described, and in the latter, not all string-theory
contributions to the vacuum energy were taken into account.

One way to seek for physically motivated models is to try to
represent them in a field theoretical language, the easiest way
being through scalar fields. If one is able to find a scalar field
counterpart for a particular $\Lambda(t)$ version, it is natural
to extend the model to other spacetimes and other gravitational
theories, like an effective high energy string cosmology, for
example. Another advantage is the possibility of quantizing the
scalar field, which can help to find a fundamental justification
for the phenomenological model. The procedure presented here works for a
real and minimally coupled scalar field and generic $\Lambda(t)$ models,
but it can be generalized for other cosmologies.

Following standard lines, the vacuum energy density and pressure
into Eqs. (\ref{eq:rho4}) and (\ref{eq:p4}) are replaced by the
corresponding scalar field expressions, that is, ${\Lambda \over
8\pi G}\to \rho_{\phi} ={\dot\phi^2 \over 2} + V(\phi)$ and
$-{\Lambda \over 8\pi G}\to p_{\phi} ={\dot\phi^2 \over 2} -
V(\phi)$. The resulting EFE equations are:

\beq \label{friedmann00} 3H^2 = {8 \pi G}\left({\dot\phi^2 \over
2} + V(\phi) + \tilde \rho \right), \eeq \beq\label{friedmannii}
3H^2 + 2\dot H = - {8\pi G}\left( {\dot\phi^2 \over 2} - V(\phi) +
\tilde p \right),
\eeq

\beq\label{conserva} \dot{\tilde\rho} +3\gamma H \tilde\rho =
-\dot\phi (\ddot\phi + 3H\dot\phi + V'(\phi )) \equiv \tilde F.
\eeq where a tilde is used in the fluid component quantities in
order to distinguish their values from their $\Lambda(t)$
counterparts. As remarked before, this is necessary because
although dynamically equivalent (that is, having the same $\geff$)
the two versions are, in principle, thermodynamically different.
It can be seen that the above equations imply that the dynamic
parameter

\beq\label{eq:geffp}
\geff = {\dot\phi^2 + \gamma\trho
\over \trho_t},
\eeq
where $\trho_t = {\dot\phi^2 \over 2} + V(\phi) + \trho$ is the total
energy density.

Now, in order to separate the scalar field contributions, it is
interesting to introduce a second dimensionless
parameter\cite{PRD2}

\beq\label{eq:x} x\equiv {\dot\phi^2 \over \dot\phi^2 +
\gamma\trho}, \eeq which may be understood as follows. In order to
evaluate how the potential energy $V$ is distributed, it is
convenient to compare the kinetic term $\dot\phi^{2}=\rho_\phi +
p_\phi$ with a quantity involving the energy of the material
component. For the mixture, one such a quantity is $\trho_t +
\tp_t = \dot\phi^2 + \gamma\trho$, which involves the redshifting
terms of the Friedmann equation. As a measure of the relative
weights of the redshifting components, $x$ indirectly quantifies
the amount of energy that the potential $V(\phi)$ is delivering
to each component along the universe evolution, and as such, $x$
is a quantity dependent on the thermodynamic conditions underlying
the decaying process of the scalar field. Perhaps not less
importantly, $x$ simplifies considerably the subsequent equations,
since each part appearing in $\trho_t$ can be rewritten in terms
of $x$ and $\geff$,

\beq \tilde \rho = {3H^2\over 8\pi G}
{\geff\over\gamma}(1-x)=\rho(1 - x), \eeq

\beq \label {eq:dotphi} \dot\phi^2 = {3H^2\over 8\pi G} \geff x,
\eeq

\beq\label{eq:V} V(H(\phi)) = {3H^2\over 8\pi G} \left[1 -\geff
\left({x\over 2} + {1-x \over \gamma}\right)\right],
\eeq
thereby allowing a direct comparison  with the related quantities of the
dynamic $\Lambda(t)$ case.

	It should be stressed that the fluid component
generated by the decaying scalar field is different from the one created by
the decaying cosmological term. Besides, the mathematical problem has
been restated in such a manner that, given $\geff$ and $x$, one
has the evolution of the ``three'' effective components, namely,
$\tilde \rho$, $\dot\phi^2$, and $V(\phi)$. Now, assuming that
both parameters are functions of time, or equivalently, of the
scale factor or of the Hubble parameter, it can be shown that $\phi$
is given by one of the following forms:
\begin{eqnarray}
\phi-\phi_I & = & \pm \sqrt{3\over 8\pi G} \int^t_{t_I}\sqrt{\geff
x} H dt,
                   \label{eq:phi(t)}\\
            & = & \pm \sqrt{3\over 8\pi G} \int^a_{a_I}\sqrt{\geff x} {da
\over a}, \label{eq:phi(a)}\\
            & = & \pm {1\over \sqrt{6\pi G}} \int^{H_I}_H \sqrt{x \over \geff}
                      {dH \over H},  \label{eq:phi(H)}
\end{eqnarray}
where (\ref{eq:ef}) was used in the last expression. In principle,
once $\phi(t)$, $\phi(a)$ or $\phi(H)$ is found, the inversion of
the resulting expression can be done to get an explicit form for
$V(\phi)$. Of course, if the expression is not directly
invertible, a parametric reconstruction of the potential can still
be obtained. As might be expected, two extreme and somewhat
trivial situations arise naturally: (i) For $x=0$ it is seen from
(\ref{eq:x}) that $\dot\phi^2=0$, thereby recovering exactly the
original $\Lambda(t)$ scenario (ii) If $x=1$ one obtains $\tilde
\rho=0$ which leads to the limit of a pure scalar field evolution,
i.e., the universe evolves with no matter, which is the situation
originally dealt by Ellis and Madsen \cite{Ellis}. Nontrivial and
physically interesting scenarios request an intermediary value of
the $x$ parameter ($0<x<1$).

Generically, for a given $\Lambda(t)$, the $\geff$
parameter is obtained by using (\ref{eq:geffL}) and $H$ is found
by solving (\ref{eq:ef}), but to compute the remaining functions,
$x$ must be specified. This can be done by imposing to the scalar
field picture the same thermodynamic conditions applied to the
$\Lambda(t)$ model, including the ``adiabatic''
particle production (see, for instance, \cite{JL}). In this case,
the scalar field thermodynamics with a dissipative term is the
same used in the last section (assuming that a classical scalar
field do not carry entropy), with two important exceptions: the
source term in the energy conservation equation

\beq\label{eq:tildeF} \tilde F = 3H\gamma\trho \left[1 -
{\geff\over\gamma} + {1\over 3H\gamma}
    \left( {\dot\geff\over\geff} - {\dot x\over 1-x}\right)\right],
\eeq and the temperature law

\beq\label{eq:EVOLTp} {\dot{\tT} \over \tT} = -3H{\gamma - 1 \over
\gamma} \left[\geff - {1\over 3H}\left({\dot\geff\over\geff}-
        {\dot x \over 1-x}\right)\right].
\eeq
Note, however, that the functional relation between the particle number and
energy source terms is exactly the same (see Eq. (\ref{adc}))

\beq \tilde\Gamma = {\tilde F \over \gamma\tilde\rho}.
\eeq

As it appears, a scalar field cosmology can be considered as
equivalent, or at least analogue to a dynamic $\Lambda$ version if
they have the same dynamics (parametrized by $\geff$) and the same
temperature evolution law. By comparing Eqs. (\ref{eq:EVOLTL}) and
(\ref{eq:EVOLTp}) one can see that this constraint is satisfied
for $x={\rm const.}$ This is quite convenient, since the integrals
(\ref{eq:phi(t)})-(\ref{eq:phi(H)}) depend only on $\geff$ under
such a condition. As one may check, a constant $x$ also implies
that

\beq \label{eq:rates} {\dot S\over S}= {\dot {\tilde S}\over
\tilde S}, \quad {\dot \rho \over \rho}= {\dot {\tilde \rho}\over
\tilde \rho}, \quad \Gamma = \tilde\Gamma,
\eeq
for any dynamics,
thereby reinforcing this interpretation of thermodynamic analogy.
In this context, $x$ is a new phenomenological parameter that must
be limited or obtained from cosmological data. This is consistent
with the fact that scalar field cosmologies have one extra degree
of freedom since there are two functions (the potential and the
kinetic term) to be determined, instead of one, $\Lambda(t)$, in
the decaying vacuum picture. The usefulness of the procedure
outlined above may be better evaluated by working on particular
models.

In order to make the basic steps of the algorithm more explicit,
it may be pedagogical to apply it in the simplest possible
situation. The most straightforward model to be considered is the
one of Freese {\em et al.} \cite{FF87} (see also \cite{JLW92,MPLA}),
which is characterized by the function \beq \Lambda (H) = 3\beta
H^2, \eeq where the dimensionless $\beta$ parameter is contained
on the interval $[0,1]$. Inserting this expression into
(\ref{eq:geffL}), one gets $\geff = \gamma (1-\beta )$.
Integrating (\ref{eq:phi(H)}) and substituting the result in
(\ref{eq:V}), one has

\beq\label{eq:expot} V(\phi) = V_Ie^{-\lambda(\phi-\phi_I)},
\eeq
where it was assumed that $\phi > \phi_I$, $\lambda = \sqrt{3\pi
G\gamma(1-\beta)/2x}$, and $V_I$ is the expression (\ref{eq:V})
with $H=H_I$. This potential was already considered in \cite{PRD2}, and
can be interpreted as a sort of ``coupled quintessence'' \cite{Amendola}.
One might also notice that the exponential potential would be obtained even
if $\beta= 0$ (no particle production, like in some quintessence models)
and $x=1$ (absence of particles, equivalent to power-law inflation, if
$\gamma < 2/3$).

There are two comments that should be made about the above
procedure. Occasionally, one may have to use the $\geff$ obtained
from (\ref{eq:geffL}) and solve (\ref{eq:ef}) first, in order to
get an explicit functional relation for $H$, and then proceed to
the subsequent steps to find $V(\phi)$. Naturally, there will be
cases in which the system of equations will note have exact
solutions, so that numeric calculations should be needed. Another
important point is that, although analogue, the scalar field
picture presented here is not an exact description of its
$\Lambda$ version. Despite having the same temporal rates and
following the same dynamics, the thermodynamic quantities have
their values parameterized by the constant $x$, if compared to
their $\Lambda$ picture counterparts. This result may have
phenomenological consequences. For instance, the temperatures of
these two pictures are related by

\beq \tilde T = (1-x)^{1\over 4}T,
\eeq
so that physical processes
like nucleosynthesis and matter-radiation decoupling happen at
different times, and phenomenological bounds on decaying $\Lambda$
cosmologies, like the nucleosynthesis bounds found in \cite{BBN},
should be reevaluated.

\section{A case study: the deflationary universe}

In previous papers \cite{L&M 94}, it was proposed a
phenomenological decaying-$\Lambda$ law that yielded a nonsingular
cosmological scenario of the deflationary type (as Barrow
\cite{Barrow} termed the pioneering Murphy's model \cite{Murphy}).
In this model, the cosmic history started from an instability of
the de Sitter spacetime in the past infinity, and, subsequently,
the universe evolved towards a slightly modified
Friedmann-Robertson-Walker (MFRW) cosmology. The solution may be
described analytically, and the transition from the de Sitter to
the MFRW phase is continuous with the decaying vacuum generating
all the matter-radiation and dark energy of the present day
universe. Broadly speaking, it resembles the warm inflationary
scenarios \cite{WI} since the particle production process occurs
along the inflationary phase and a reheating phase is not
necessary. In addition, the maximum allowed value for the vacuum
energy density may be larger than its present value by about $120$
orders of magnitude, as theoretically suggested.

The above discussion shows that the important question here is how
the vacuum energy density decays in the course of the expansion.
Note that for a two component fluid it is natural to introduce the
parameter:
\beq\label{eq:beta} \beta \equiv {\rho_v
\over \rho + \rho_v} = {\rho_v\over \rho_t}.
\eeq
This means that
$\rho_v = 3\beta H^{2}/8\pi G$ or equivalently $\Lambda (H) =
3\beta H^{2}$ as claimed by the authors of reference \cite{JLW92}
using dimensional arguments (see also \cite{FF87,MPLA}). The first
case analysed was $\beta$ constant \cite{FF87}. However, it has
been shown that such a scenario does not started from a de Sitter
universe \cite{JLW92} so that a deflationary scenario is absent.
In this way, a description of an earlier inflationary period with
no matter (de Sitter) characterized by a definite time scale
${H_I}^{-1}$ is possible only if one consider a time-dependent
$\beta$ parameter, or equivalently, higher order terms of $H$ in
the expansion of $\Lambda(H)$. In order to show the plausibility
of the expression proposed in \cite{L&M 94}, consider the
latter approach through the expansion

\beq \label{defLambda}
\Lambda (H) = 3\beta H^{2} + \delta
{H}^{3},
\eeq
where $\delta$ is a dimensional parameter, $[\delta]
= ({\rm time})^{-1}$, which must be fixed by the time scale of
deflation. As one may check, the de Sitter condition,
$max[\Lambda]=3{H_{I}}^{2}$, implies that $\delta = 3(1 -
\beta)H_I^{-1}$. Inserting this result into (\ref{defLambda}) the
phenomenological law considered in Ref. \cite{L&M 94} is obtained,
namely:

\beq\label{eq:def} \Lambda (H) = 3\beta H^{2} + 3(1 -
\beta){{H}^{3}\over H_I}, \eeq or still, \beq\label{eq:def1}
\rho_v = \beta \rho_t \left( 1 + {(1 - \beta) \over
\beta}{{H}\over H_I}\right),
\eeq
and comparing (\ref{eq:def1})
with (\ref{eq:beta}), it is seen that the fractional vacuum to
total density ratio parameter is now a time-dependent quantity.

The arbitrary time scale $H_{I}^{-1}$ characterizes the initial de
Sitter phase, and, together with the $\beta$ parameter, is
presumably given by fundamental physics. At late times ($H \ll
H_{I}$), the second term on the right hand side of (\ref{eq:def})
can be neglected. This means that the $\beta$ coefficient
measures the extent to what the model departs from the
standard flat FRW cosmology at late stages. This model may also be
viewed as an early phase of the decaying $\Lambda$ scenario
originally proposed in \cite{FF87}.

Now, inserting the expression of $\Lambda (H)$ into
(\ref{eq:geffL}) the following expression for the $\geff$ is
obtained
\beq \label{eq:geffdefla} \geff = \gamma
(1-\beta)\left(1-{H \over H_I}\right).
\eeq
Note that for $H=H_I$
this equation describes the de Sitter space-time and gives the
maximum value for the cosmological term, which corresponds to the
value of $\Lambda$ for the unstable de Sitter phase. For $H \ll
H_I$ the model behaves like a MFRW model modified by the
$\beta$ parameter. Following \cite{L&M 94}, the transition from de
Sitter to the MFRW phase is exactly described as \beq
\label{eq:H(a)} H = {H_I \over 1 + {H_I\over H_0} \left(a \over
a_{0}\right)^{3\gamma(1-\beta)\over 2}}, \eeq where the subscript
$0$ refers to present time quantities. Integrating the above
equation, one gets

\beq \label{eq:t(a)} t = t_e + {1\over H_I}\left[\ln \left({a\over
a_e}\right) + {2\over 3\gamma (1-\beta )}{H_I\over H_0}
\left({a\over a_0}\right)^{3\gamma (1-\beta)\over 2}\right],
\eeq
where ``$e$'' denotes the end of inflation and it was assumed that
$H_I \gg H_0$ and $a_0\gg a_e$. Recently, the above solution has
also been discussed in a similar context by Gunzig et al.
\cite{GMN} for the particular case with $\gamma = 4/3$ and $\beta
= 0$.

As one may check, the matter-radiation energy density is given by

\beq \label{eq:rho(H)4} \rho = {3(1-\beta) \over 8\pi G}H^2
\left[1-\left({H\over H_I}\right)\right],
\eeq
and it can be shown
that its maximum value is
\beq \label{eq:rhomax} \rho_m =
{(1-\beta) \over 18\pi G}H_I^2.
\eeq
Therefore, the fluid
component starts with a zero value for $H=H_I$, grows until
$\rho_m$ (for $H_m = (2/3)H_I$) and decreases throughout a MFRW
phase which dynamically resembles some recent dark energy models
\cite{quintessence} with tracking solutions, but presenting
particle production. The Hubble parameter at the end of inflation
($\ddot a=0$) can be written as

\beq \label{eq:HeHi} H_e = H_I \left( 1 - {2\over
3\gamma(1-\beta)}\right).
\eeq

Now, since for the radiation dominated phase, $\rho \propto T^4$,
and from (\ref{eq:rhomax}), the maximum temperature reached by the
gas is

\beq\label{eq:tmax} T_m = \left({30\rho_m\over \pi
g_*}\right)^{1\over 4},
\eeq
where $g_*$ is the number of spin
degrees of freedom of the fluid components. From (\ref{eq:rhomax})
and (\ref{eq:HeHi})

\beq\label{eq:HeHm}
  H_e ={3\over 2} H_m \left( 1 - {2\over 3\gamma(1-\beta)}\right),
\eeq
so that $H_e < H_m$ for any values of $\gamma$ and $\beta$
except $\gamma = 2$ and $\beta = 0$. Generically, this means that
the universe attains its maximum temperature before the end of
inflation. In this connection, an interesting and careful
thermodynamic analysis of deflationary models including other
processes than vacuum decay can also be found in Refs.
\cite{Zimdahl,ZimBal}.

As argued in the last section, a scalar field equivalent to the
model above outlined can be obtained by using a constant $x$ and
the deflationary expression for $\geff$ given by Eq.
(\ref{eq:geffdefla}). In such a case, the following analytical
expression for the scalar field potential is found

\beq \label{eq:Vphi} V(\phi)={3H^2_I\over 8\pi G} {1- \epsilon
\tanh^2 [\lambda(\phi - \phi_I)]\over \cosh^{4}\left[\lambda(\phi
- \phi_I)\right]} .
\eeq
for $\phi > \phi_I$, $\epsilon =
(1-\beta)[1 - x(1 - \gamma/2)]$ and $\lambda = \sqrt{ 6\pi G\gamma
(1-\beta)\over 4x}$. As one should expect, the above potential goes
asymptotically to the exponential potential (\ref{eq:expot}). Furthermore,
it was obtained for a constant $x$, but if the required thermodynamic
equivalence is relaxed, $x$ can be a variable quantity, and
different potentials can be found for the same deflationary
behavior. Even the exponential potential would be obtained if, for
instance, $x\propto \geff$. Of course, as can be seen from
(\ref{eq:tildeF}) or (\ref{eq:EVOLTp}), in order to get a
time-varying $x$ one would have to provide appropriate
thermodynamic conditions for the system.

Potentials with the funtional form of (\ref{eq:Vphi}) were also
found by Maartens {\em et al.} \cite{Maartensetal} and by Zimdahl
\cite{Zimdahl}, but their scenarios have important differences
with respect to the deflationary model studied here. In the
first case, the authors considered $x=1$ (no matter or radiation)
and represented only the deflationary dynamics, with a transition
from a de Sitter for a radiation-like dynamics driven solely by
the interchange of energy between the $\phi$ potential and its
kinetic term. Their potential is a particular case of the one
presented here, with $\gamma = 4/3$, $x=1$ and $\beta=0$. Zimdahl
obtains the same potential of Maartens {\em et al.} (and the same
values for $\gamma$, $\beta$ and $x$), but in a different
thermodynamic context, so that it cannot be considered as a case
of (\ref{eq:Vphi}). Another important difference is that the
particle creation rate $\tilde\Gamma$ in Zimdahl's work is not the
same one of the deflationary $\Lambda(t)$ picture treated here,
thereby violating one the equalities (\ref{eq:rates}). Besides,
the initial value for the fluid energy density in his picture is
not zero, so that there is not a thermodynamic  equivalence between the
$\Lambda(t)$ and scalar field counterparts in the sense above discussed.

Another interesting feature of such models is that the scalar
field and the fluid are thermally coupled during the inflationary
era so that they can be regarded as instances of warm inflationary
scenarios \cite{WI}. The connection between the two approaches is
just the function $\tilde F$. In warm inflation, this thermal
coupling is represented by a dissipative term
$\Gamma_{\phi}\dot\phi^2$ in the energy conservation equation. As
usually the adiabatic condition or, alternatively, the perfect
fluid thermodynamic relation ($\trho \propto \tilde T^4$) is
assumed, one may write $\Gamma_{\phi}$ in terms of the particle
creation rate $\tilde\Gamma$:

\beq \Gamma_{\phi} = {1-x\over x}\tilde\Gamma.
\eeq
For a constant
$x$ (that is, $\Gamma = \tilde\Gamma$), the above equation allows
one to translate the results of a given dynamical $\Lambda(t)$
model to its warm inflationary counterpart. Furthermore, in such a
case $\Gamma_{\phi}$ has the direct interpretation of a particle
production rate (in a comoving volume). Since warm
inflation does not have an uncontroversial derivation from first
principles (see, for example \cite{BGR}), the methods employed
here might also be useful in the search for a more fundamental
justification of warm scenarios.

\section{Conclusion}

A procedure to write scalar field versions of decaying vacuum
cosmologies has been proposed. The switching between the two
pictures depends basically on a pair of parameters ($\geff,
x$) conveniently chosen. The first one, $\geff$, is
responsible for the dynamic equivalence while the $x$ parameter,
measuring the fraction of energy carried by the redshifting
components of the universe, can be related to the thermodynamic behavior.
It has been shown that the two pictures can be made dynamically and
thermodynamically equivalent in the sense that the cosmological
quantities may have the same evolution laws. In a worked example
(the deflationary cosmology), the time independent $x$ parameter
gave rise to an exponential potential for the scalar field, but
other potentials can be found if $x$ is allowed to vary in the
course of the evolution.

It is worth mentioning that the set of equations presented in
Section III is not necessarily limited to the search for scalar
field counterparts of decaying vacuum scenarios. In principle,
the procedure discussed here opens the possibility of scalar field
versions involving other phenomenological descriptions, like
bulk viscosity and matter creation cosmologies.
In this connection, it is also interesting to investigate how the
causal thermodynamic approach (or second order theories) \cite{IP}
constrains the $x$ parameter.

Finally, since a time-varying $x$ can be used together with
$\geff$ as free parameters, this approach can be applied to
solve ``inverse problems'' by using present and future
observational data to reconstruct the potential $V(\phi)$ either
by considering particle production \cite{Dalal} or universes
containing a perfect fluid plus a decoupled scalar field
\cite{Saini}. Further investigations in this direction are in course.

{\bf Acknowledgments:} The authors thank Ana Helena de Campos for
reading the manuscript and Luca Amendola for a helpful discussion.
JMFM thanks FAPESP for supporting this work. JASL is also
partially supported by CNPq.

\end{document}